\documentclass[12pt]{article}
\usepackage{epsfig}
\title{Lorentz-Invariant Interpretation of Noncommutative Space-Time - global version. }
\author{Piotr Kosi\'nski\thanks{supported by U\L $ $\ grant no. 690}, Pawe\l $$\ Ma\'slanka$^*$
\\Department of Theoretical Physics II \\
University of {\L}\'od\'z \\
Pomorska 149/153, 90 - 236 {\L}\'od\'z/Poland.}
\date{}

\begin{document}
\maketitle
\begin{abstract}
The global version of the quantum symmetry defined by Chaichian et al (hep-th/0408069) is constructed.
\end{abstract}

\newpage

In  the very recent interesting paper Chaichian and al. \cite{b1} proposed a new interpretation of the symmetry 
of noncommutative space-time defined by the commutation relations:
\begin{eqnarray}
[x_{\mu},\;x_{\nu} ]=i\Theta_{\mu \nu}, \label{w1}
\end{eqnarray}
where $\Theta_{\mu \nu}$\ is a constant antisymmetric matrix. According to the standard wisdom the relations (\ref{w1})
 break the Lorentz symmetry
down to the stability subgroup of $\Theta_{\mu \nu}$. In spite of that all fundamental issues of the noncommutative quantum 
field theory (NCQFT) are discussed in fully covariant approach using the representations of Poincare group. 
The reason for that might be that NCQFT emerges as specific limit of fully symmetric theory.

On the other hand one can pose the question whether the noncommutative space-time admits as large symmetry 
as its commutative counterpart provided the 
symmetry is understood in the wider sense (for example, as a symmetry in the sense of the quantum group theory). This
 is important because if one tries to base the theory on the stability subgroup of $\Theta_{\mu \nu}$\ 
one is faced at once with deep problems \cite{b2} (for example, why the multiplets of stability subgroup 
are organized in such a way as to form the complete multiplets of the whole group).
The problem can be posed in quite general terms. Given a theory based on some  
symmetry group broken explicitly to its subgroup it is usually not sufficient to use the properties of this subgroup. 
Some questions can be answered only within the framework of the initial symmetry in spite of the fact that it is broken,
i.e. it is formally no longer a symmetry.

 The solution to this dilemma might be as follows. Assume the original symmetry group is broken by some 
additional conditions imposed (like Poincare group being broken by a specific choice of $\Theta_{\mu \nu}$). 
Then it appears that some properties of 
the system can be explained in terms of the residual symmetry while in order to understand other properties one 
has to appeal to the initial symmetry. Assume further that we 
have found a quantum symmetry as large as the initial classical one. It provides a deformation of the classical symmetry, 
the parameter of deformation being determined by the strength of symmetry breaking. Quantum symmetry is a more
 general notion
and, therefore, one can expect its consequences are weaker. Ideally, we can hope that the quantum symmetry implies
 some conclusions of
initial classical symmetry (e.g. the classification of multiplets) survive while other (modified by symmetry breaking) do not.

It has been shown in ref. \cite{b1} that the quantum symmetry of noncommutative space-time defined by eq. (\ref{w1}) is
a twisted Poincare algebra, the twist element being an abelian twist \cite{b3}:
\begin{eqnarray}
F= exp(\frac{i}{2}\Theta^{\mu \nu}P_{\mu}\otimes P_{\nu})\label{w2}
\end{eqnarray}
Twisting the Poincare algebra provides an infinitesimal form of quantum symmetry of noncommutative space-time. The first 
step in checking whether the above sketched scenario works in NCQFT is to analyse the mathematical structure of the quantum symmetry
found in ref. \cite{b1}. In the present note we give the global version of twisted Poincare symmetry of Chaichian et al.

 Our starting point is
the matrix form of Poincare transformations. Namely, we consider $5\times 5$\  matrices $T^a_{\;b},\;a,\;b=0,\;1,\;...,\;4$,
of the form
\begin{eqnarray}
T=\left[\matrix{\Lambda ^{\mu}_{\;\;\nu}&\mid &a^{\mu}\cr----&\mid & ----\cr 0&\mid &1\cr}\right];\label{w3}
\end{eqnarray}
here $\mu,\;\;\nu =0,\;...,\;3,\;\Lambda^{\mu}_{\;\;\nu}$\ is Lorentz matrix while $a^{\mu}$\ denotes translation.
 The composition law for 
Poincare group can be now written as
\begin{eqnarray}
\Delta T^a_{\;\;b}=T^a_{\;\;c}\otimes T^c_{\;\;b}\label{w4}
\end{eqnarray}
We take eq. (\ref{w4}) as the definition of coproduct of our quantum Poincare group. In order to find the 
algebraic structure one can use the 
FRT relation \cite{b4}
\begin{eqnarray}
RTT=TTR \label{w5}
\end{eqnarray}
where $R$\ is the universal $R$-matrix in the representation determined by $T$. Now, the $R$-matrix for a given
 twist $F$\ of 
classical group reads \cite{b3}
\begin{eqnarray}
R=F_{21}F^{-1} \label{w6}
\end{eqnarray}

In our case $F$\ is given by eq. (\ref{w2}) while $P_{\mu}$\ can be computed from (\ref{w3}). The calculations are greatly
simplified by the fact that $P_{\mu}$\ are nilpotent matrices.

 Skipping the details we present the final result. 
The quantum 
Poincare group dual to the algebra considered in ref. \cite{b1} is defined by the following relations
\begin{eqnarray}
&&\Delta \Lambda^{\mu}_{\;\;\nu}=\Lambda^{\mu}_{\;\;\alpha}\otimes \Lambda^{\alpha}_{\;\;\nu} \nonumber \\
&&\Delta a^{\mu}=\Lambda^{\mu}_{\;\;\alpha}\otimes a^{\alpha}+a^{\mu}\otimes 1 \nonumber \\
&&\varepsilon (\Lambda^{\mu}_{\;\;\nu})=\delta^{\mu}_{\;\;\nu} \label{w7} \\
&&\varepsilon (a^{\mu})=0\nonumber \\
&&S(\Lambda^{\mu}_{\;\;\nu})=\Lambda _{\nu}^{\;\;\mu} \nonumber \\
&&S(a^{\mu})=-\Lambda_{\alpha}^{\;\mu}a^{\alpha}\nonumber \\
&&[\Lambda^{\mu}_{\;\;\nu},\cdot]=0\nonumber \\
&&[a^{\mu},\;a^{\nu}]=-i\Theta^{\rho \sigma}(\Lambda^{\mu}_{\;\;\rho}\Lambda^{\nu}_{\;\;\sigma}-\delta^{\mu}_{\;\;\rho}
\delta^{\nu}_{\;\;\sigma})\nonumber
\end{eqnarray}
and $*$-involution is defined by
\begin{eqnarray}
&&(a^{\mu})^*=a^{\mu}\nonumber \\
&&(\Lambda^{\mu}_{\;\;\nu})^*=\Lambda^{\mu}_{\;\;\nu} \label{w8}
\end{eqnarray}
One can check by straightforward calculations that the above structure is consistent and defines $*$-Hopf algebra.
Note that, contrary to the $\kappa$-Poincare case \cite{b5}, the translations do not form a subalgebra. 
In spite of that, one can define the action 
of our Poincare group on quantum space-time defined by eqs. (\ref{w1}) by
\begin{eqnarray}
x^{\mu}\rightarrow \Lambda^{\mu}_{\;\;\nu}\otimes x^{\nu}+a^{\mu}\otimes I \label{w9}
\end{eqnarray}
It is easy to check that this (co-) action is well defined and consistent with the commutation rules (\ref{w1}). 

Having defined the deformed Poincare group and its action on quantum Minkowski space (eqs. (\ref{w7})-(\ref{w9})) 
one can follow
the standard rules of quantum group theory \cite{b6} to find the representations, to classify the differential
 calculi on quantum group and representation space, etc. This should provide the tools for constructing a quantum group
 invariant dynamics and to see whether one can explain the structure of NCQFT without appealing to larger structures.

{\bf Note added } \\
After submitting this paper to hep archive we have learned from R. Oeckl that our result is contained in his paper in Nucl.
Phys.B 581, (2000), 559.


\begin{thebibliography}{99}
\bibitem{b1}
M. Chaichian, P. P. Kulish, K. Nishijima and A. Tureanu, "On a Lorentz-Invariant Interpretation of Noncommutative
Space-Time and Its Implications on Noncommutative QFT", hep-th/0408069
\bibitem{b2}
L. Alvarez-Gaume, J. L. F. Barbon and R. Zwicky, JHEP 0105 (2001), 057, (hep-th/0103069) \\
Y. Liao and K. Sibold, Phys. Lett. \underline{B549} (2002), 352, (hep-th/0209221)\\
L. Alvarez-Gaume and M. A. Vazquez-Mozo, Nucl. Phys \underline{B668} (2003), 293 (hep-th/0305093)\\
M. Chaichian, M. N. Mnatsakanova, K. Nishijima, A. Tureanu and Yu. S. Vernov "Towards an Axiomatic Formulation of Noncommutative Quantum Field Theory", hep-th/0402212\\
M. Chaichian and A. Turneanu "Jost-Lehmann-Dyson Representation and Froissart-Martin Bound in Quantum Field Theory on Noncommutative Space-Time", hep-th/0403032
\bibitem{b3}
N. Yu. Reshetikhin, Lett. Math. Phys. \underline{20} (1990), 331.
\bibitem{b4}
L. D. Fadeev, N. Yu. Reshetikhin and L. A. Takhtajan, Algebra i Analiz {\bf 1}, (1989), 178
\bibitem{b5}
S. Zakrzewski, J. Phys. \underline{A27}, (1994), 2075\\
P. Kosinski, P. Maslanka, in:"From Quantum Field Theory to Quantum Groups", e.d. B. Jancewicz and J. Sobczyk, World Sc., 1996, p. 41\\
P. Kosinski, J. Lukierski, P. Maslanka, Phys. Rev. \underline{D62} (2000), 025004.
\bibitem{b6}
see for example, V. Chari and A. Pressley, "A Guide to Quantum Groups", Cambridge University Press, 1994\\
M. Chaichian and V. Demichev, "Introduction to Quantum Groups", World Sc. 1996\\
S. L. Woronowicz, Comm. Math. Phys. \underline{111}, (1987), 613\\
S. L. Woronowicz, Comm. Math. Phys. \underline{122}, (1989), 125
\end{thebibliography}
\end{document}